\documentclass{article}
\begin{document}
\title{On the Quantization of the Monoatomic Ideal Gas}

\author{E. Fermi \\ \\ translated by \\ A. Zannoni \\ \it{Institute for Physical Science
and Technology}, \\ \it{University of Maryland at College Park},\\ \it{College Park,
Maryland 20742}\\ and \\ \it{Electron and Optical Physics Division},\\ \it{National
Institute of Standards and Technology},\\ \it{Gaithersburg, Maryland 20899}}
\date{ }

\maketitle
 \noindent{\bf{\Large Introduction}}\\

A recent experiment \cite{JILA} has brought a nearly-ideal gas of Fermi-Dirac particles ($^{40}$K atoms) to the condition of {\it quantum degeneracy}, in which the symmetry of the many-particle wavefunction has a dominant effect on the equation of state of the gas.

In the above-mentioned experiment, the gas atoms were confined by a magnetic field, whose effect is expressed by the harmonic potential,
\begin{eqnarray}
\nonumber V(x,y,z) = \frac{M}{2}\left(w^2_{r}(x^2+y^2)+w^2_{z} z^2\right)
\end{eqnarray}
where $x, y, z$ are Cartesian particle coordinates, $M$ is the particle mass, $\omega_r = 2\pi\, 137$ Hz, and  $\omega_z = 2\pi\, 19.5$ Hz.
This situation is close to that portrayed in the original model used by
Fermi to elucidate the effects of quantum degeneracy on the
equation of state of an ideal gas, in the paper translated here.\cite{Lincei}
Fermi confined the gas with an isotropic harmonic
oscillator potential ($\omega_r = \omega_z$).  This approach
contrasts with that of most modern textbook treatments, which put the
particles in a volume of constant potential, subject to either hard-wall or
periodic boundary conditions, as are appropriate to the treatment of
extended, homogeneous systems, {\it e.g.} electrons in a metal.  The utility
of such an approach was recognized by Fermi, but he adopted the harmonic
confining potential so as to allow the invocation of the Bohr-Sommerfeld quantization rule in its simplest form.

In passing to the thermodynamic limit, the essential physics of the ideal Fermi-
Dirac gas emerges independently of the choice of confining potential.  Thus
Fermi's treatment can be enjoyed without reference to recent developments.  However, since some details of his original derivation have
just come to have specific relevance to experiments, it seems appropriate to
make them more accessible to an English-speaking audience.

The translation provided here is based on the Italian original,\cite{Lincei} "Sulla
quantizzazione del gas perfetto monoatomico," as reprinted in the first volume of
Fermi's collected papers.\cite{Midway} A longer, German-language paper, presenting the
argument in greater detail, was subsequently published by Fermi in {\it Zeitschrift
f\"ur Physik}.\cite{Z} That paper is also reprinted in Ref. (\cite{Midway}).

I thank Charles Clark and Maria Colarusso for comments on the manuscript.\\
\\

\noindent{\bf{\Large On the Quantization of the Monoatomic Ideal Gas}\\
{\small ``Rend. Lincei'',3,145-149 (1926),\\
presented by the Associate A. Garbasso at the meeting of 7 February 1926}}\\

1. According to classical thermodynamics, the specific heat at constant volume  of a
monoatomic ideal gas (referred to a single molecule) is given by $c=3k/2$. It is clear
however that if we also want to admit the validity of the Nernst principle for an
ideal gas, we need to consider the previous expression for $c$ as only an
approximation for high temperatures, and in reality $c$ goes to zero for $T=0$, so
that we can extend down to absolute zero the integral that expresses the value of the
entropy without leaving the constant undetermined. To realize how this kind of
variation in $c$ can take place, it is necessary to admit that also the motions of the
ideal gas have to be quantized. It is understandable also that such quantization will
affect not only the amount of energy of the gas, but also its equation of state,
giving the so-called degeneracy phenomena of the ideal gas for low temperatures.

The purpose of this work is to present a method that makes the quantization of the ideal
gas possible, and which is, we believe, as independent as possible from unjustified
hypotheses concerning the statistical behavior of the molecules of the gas.

Recently many attempts have been made to establish the equation of state for an ideal
gas.\cite{Al} The formulae given by the various authors and ours differ from each
other and from the classical equation of state, only for very low temperatures and
very high densities; unfortunately these are the same conditions for which the
differences between the laws of real gases and ideal gases are most important. Since
under conveniently attainable experimental conditions, the deviations from the
equation of state $pV=kT$ caused by the degeneracy of the gas, even if not negligible,
are always considerably smaller than those due to the fact that the gas is real and
not ideal, so that up to the present the former have been masked by the latter; it
cannot be excluded that, with an improved knowledge of the forces acting between the
molecules of a real gas, it might be possible, more or less in the near future, to
separate the two deviations, so as to decide experimentally between the different
theories of the degeneration of the ideal gas.

2. In order to be able to perform the quantization of the motion of the molecules of
an ideal gas it is necessary to put ourselves in a position to apply Sommerfeld's
rules to their motion: this can clearly be done in an infinite number of ways which,
of course, all give the same result. For example, we can suppose the gas to be
enclosed in a parallelepiped receptacle with elastic walls, quantizing the triply
periodic motion of the molecule that bounces between the six faces of the receptacle;
or, more generally, we can apply to the molecules a suitable system of forces such
that their motion becomes periodic and can then be quantized. The hypothesis that the
gas is ideal allows us to neglect in all these cases the forces acting between the
molecules, so that the mechanical motion of each one of these takes place just as if
the others did not exist. It is possible however to realize that the simple
quantization, with Sommerfeld's rules, of the motion of the molecules considered as
completely independent from each other is not sufficient to get correct results;
because, even if in this way the specific heat goes to zero for $T =0$, we find that
its value depends not only on the temperature and density, but also on the total
amount of gas, and goes, for every temperature, to the limit $3k/2$ when, while
keeping the density constant, the total amount of gas goes to infinity. Thus it seems
necessary to admit that we must add some complements to Sommerfeld's rules, in the
case of systems, like ours, in which the elements are not distinguishable from each
other.\cite{Nuovo}

To have a hint of what is the most plausible hypothesis that
we can make, it is convenient for us to examine how things work for other systems
that, like our ideal gas, have indistinguishable elements; and precisely, we want to
examine the behavior of atoms heavier than hydrogen, all containing more than one
electron. If we consider the deepest parts of a heavy atom, we find conditions such
that the forces that act on the electrons are very small compared with those created
by the nucleus. In these circumstances the pure and simple application of Sommerfeld's
rules would lead us to predict that, in the normal state of the atom, a considerable
number of electrons should be found in an orbit of total quantum 1. In reality we
observe that the K ring is already saturated when it contains two electrons, and in
the same way the L ring is saturated when it contains 8 electrons, etc. This fact was
interpreted by Stoner,\cite{Stoner} and even more precisely by Pauli,\cite{Pauli} in
the following way: let us characterize a possible electronic orbit in a complex atom
with 4 quantum numbers; $n$,$k$,$j$,$m$ that have respectively the meaning of total
quantum, azimuthal quantum, internal quantum and magnetic quantum. Given the
inequalities that those four numbers have to satisfy, it is found that for $n=1$,
there exist only two triples of values $k,j,m$; for $n=2$, there exist 8 triples of
values, etc. To realize this fact, it is sufficient to assume that in the atom there
can not be two electrons with the orbits described by the same quantum numbers; in
other words it is required to admit that an electronic orbit is already "occupied" when it
contains only one electron.

3. We now want to find out if such a hypothesis can give good results also in the
problem of the quantization of the ideal gas:  we shall thus assume that in our gas
there can be at the most one molecule, whose motion is characterized by certain
quantum numbers, and we shall show that this hypothesis leads us to a perfectly
consistent theory for the quantization of the ideal gas that, in particular, accounts
for the expected decrease of the specific heat for low temperatures, and that yields
the exact value for the constant of the entropy of the ideal gas.

Reserving publication of the mathematical details of this theory to another occasion, we limit
ourselves in this Note to showing the principles of the method and the results.

First of all we have to subject our gas to such conditions that the motion of its molecules
can be quantized. As we have seen this can be done in an infinite number of ways; however as the result is independent of the specific way chosen, we shall choose the
one that makes the calculations easy; and precisely we shall assume that an attractive
force toward a fixed point $O$ acts on our molecules, with strength proportional to
the distance $r$ of the molecule from $O$; so that each molecule will become a single
spatial harmonic oscillator, with a frequency that we call $\nu$. The orbit of the
molecule will be characterized by its three quantum numbers $s_1$,$s_2$,$s_3$, that
are related to its energy through the relation

\begin{equation}
w=h \nu (s_{1}+s_{2}+s_{3})=h\nu s
\end{equation}

The energy of a molecule can thus take all the multiple integer values of $h\nu$, and
the value $s h \nu$ can be taken in $Q_s=\frac{1}{2}(s+1)(s+2)$ modes.

The zero energy can thus be realized only in one way, the energy $h\nu$ only in 3 ways, the
energy $2h\nu$ in 6 ways, etc. To realize the consequences of our hypothesis, that given quantum numbers can not correspond to more than one molecule, let us consider the
limiting case of having N molecules at the absolute zero. At this temperature the gas
has to be in the state of minimum energy. Thus, if there was no restriction on the
number of molecules that can have a certain energy, all the molecules would be in the
state of zero energy, and all the three quantum numbers of each of them would be zero.
Instead, according to our hypothesis, it is not possible to have more than one
molecule with all the three quantum numbers equal to zero; so if $N=1$, the only molecule will occupy the place with zero energy, if instead  $N=4$, one of the
molecules will occupy the place with zero energy, and the other three the places with
energy $h\nu$; if $N=10$, one of the molecules will occupy the place of zero energy,
three will occupy the three places of energy $h\nu$, and the remaining six the six
places of energy $2h \nu$, etc.

Let us suppose now that we have to distribute among
our  $N$ molecules the total energy $W=Eh\nu$ ($E$= integer number); and let us label
with $N_s\leq Q_s$ the number of molecules of energy $sh\nu$. It is easy to find
that the most probable values of $N_s$ are

\begin{equation}
\label{2}N_s=\frac{\alpha Q_s}{e^{\beta s}+\alpha}
\end{equation}

\noindent where $\alpha$ and $\beta$ are some constants dependent on $W$ and $N$. To
find the relation between these constants and the temperature, we observe that,
because of  the effect of the attraction toward $O$, the density of our gas will be a
function of $r$, that must go to zero for $r=\infty$. So, for $r=\infty$ the
degeneracy phenomena must cease, and in particular the distribution of the velocities,
easily obtainable from $(2)$, must become Maxwell's law. It is thus found that it has
to be that

\begin{equation}
\beta=\frac{h\nu}{k T}
\end{equation}

Now we are able to find from (\ref{2}) the function $n(L) d L$, that represents, for a
fixed value of $r$, the density of  molecules with an energy that ranges from $L$ to
$L+dL$ (Analogous to  Maxwell's law), and from that we can find the mean kinetic
energy $\bar L$ of the molecules at distance $r$, that is a function not only of the
temperature, but also of the density $n$.  It is found precisely that

\begin{equation}
\label{4}\bar L=\frac{3h^2n^{\frac{2}{3}}}{4\pi m}P(\frac{2\pi m k T}{h^2
n^{\frac{2}{3}}})
\end{equation}

Here we have indicated by $P(x)$ a function, with a fairly complicated analytic
expression, whose evaluation is possible when $x$ is very large or very small using the
asymptotic formulae

\begin{eqnarray}
\nonumber P(x)=x(1+2^{-\frac{5}{2}}x^{-\frac{3}{2}}+.....)\\
P(x)=\frac{1}{5}\left({\frac{9\pi}{2}}\right)^{\frac{3}{2}}\left\{1+\frac{5}{9}\left({\frac{4\pi^4}{3}}\right)^{
\frac{3}{2}}x^2+.....\right\} \label{5}
\end{eqnarray}

To deduce from (\ref{4}) the equation of state we apply the virial relation. We find
that the pressure is given by

\begin{equation}
p=\frac{2}{3}n \bar L =\frac{h^2n^{\frac{5}{3}}}{2\pi m}P(\frac{2\pi m k
T}{h^2n^{\frac{2}{3}}})
\end{equation}

In  the limit of high temperatures, that is for small degeneracy, the equation of
state thus takes the following form

\begin{equation}
p=nkT\left(1+\frac{1}{16}\frac{h^3n}{(\pi m k T)^{\frac{3}{2}}}+....\right)
\end{equation}

The pressure is therefore greater than that predicted from the classical equation of
state. For an ideal gas of the atomic weight of helium, at a temperature of $5^{o}$
absolute, and at a pressure of 10 atmospheres the difference would be of 15\%. From
(\ref{4}) and (\ref{5}) we can also deduce the expression of the specific heat for low
temperatures. It is found that

\begin{equation}
c_v=\left({\frac{16\pi^8}{9}}\right)^{\frac{1}{3}}\frac{mk^2}{h^2n^{\frac{2}{3}}}T+.....
\end{equation}

In the same way we can find the absolute value of the entropy. Performing the
calculation we find, for high temperatures

\begin{equation}
S=n\int_0^T \frac{1}{T}d\bar L=n\left(\frac{5}{2}\log{T}-\log{p}+\log{\frac{(2\pi
m)^{\frac{3}{2}}k^{\frac{5}{2}}e^{\frac{5}{2}}}{h^3}}\right)
\end{equation}

\noindent which coincides with the value of the entropy given by Tetrode and by
Stern.\cite{mynote}


\begin{thebibliography}{99}
\bibitem{mynote} Eq.(9) is reproduced here exactly from ref. 3; it is missing a factor
of $k$ on the right hand side.
\end{thebibliography}
\end{document}